# Calcul à la rupture en présence d’un écoulement à surface libre : construction de champs de pression approchés

Alain Corfdir

*CERMES, ENPC-LCPC, Institut Navier, 6 et 8 avenue Blaise Pascal, Cité Descartes, Champs sur Marne, 77455 Marne-la-Vallée cedex 2, France*

Résumé

Nous étudions la stabilité d’un milieu poreux soumis à un écoulement permanent à surface libre. Sous certaines hypothèses, il est possible d’utiliser la méthode cinématique du calcul à la rupture en utilisant un champ de pression approché par défaut. Nous nous intéressons ici à la recherche de tels champs de pression dont on puisse assurer qu’ils sont partout inférieurs à la solution exacte sans connaître cette solution exacte. Nous montrons que l’on peut utiliser comme champs approchés des champs satisfaisant des conditions relâchées par rapport au problème réel, sous réserve de l’unicité de la solution d’un problème faible associé. On traite l’exemple d’un barrage à parois verticales.

Abstract

**Yield design for porous media subjected to unconfined flow: construction of approximate pressure fields**. We consider the stability of a porous medium submitted to a steady-state flow with free-boundary. Assuming some hypotheses, it is possible to implement the kinematic method by using an approximate pressure field bounding the true pressure field from below. We are interested in finding such approximate pressure fields and in proving that they bound the true pressure field from below without knowing the true pressure field. We use fields which are solutions of a problem with relaxed conditions with regard to the real problem. Under a uniqueness condition of the solution of a weak formulation of the problem, such fields are lower bounds for the true pressure field. Finally, we give the example of a vertical dam.

borne supérieure

*Abridged English version*

In this paper, we present how to use the kinematic method of yield design [1, 2] in the case of a porous medium subjected to a flow with free surface, without evaluating the exact solution of the pressure field. The method presented here is based on the use of approximate pressure fields [3]. To implement this method, it is necessary to find approximate pressure fields bounding the true pressure field from below. We consider conditions allowing to make sure, under uniqueness assumption, that an approximate pressure field is indeed a lower bound for the true pressure field.

We recall the results of [3]. Tensile stresses are counted positively. We assume that the domain of resistance G' is defined in terms of effective stresses and satisfies (1). We denote by p the pressure field and by $\chi$ the characteristic function of the part A of the medium where p>0. Also, p' denotes an approximate pressure field and $\chi'$ the characteristic function of the part of the medium where p'>0. We assume that $p \geq p'$ (and so $\chi \geq \chi'$). We denote by $\gamma_{sat}$ the specific weight of the saturated medium and we assume that the specific weight is $\gamma_d$ in the rest of the medium, neglecting so the transition between the dry zone and the saturated zone. We consider only velocity fields satisfying condition (2). Finally, a necessary condition for stability (3) is obtained, using only the approximate pressure field.

We consider only steady-state flows. The boundary of $\Omega$ is comprised of three parts: $S_1$, $S_2$ and $S_3$ (Fig. 1). Boundary conditions are: outwards volume flow $q_d$ on $S_1$, pressure $p_d$ on $S_2 \cup S_3$. The boundary of the saturated zone A is comprised of the part of $S_1$ where p>0, of $S_3$, $\Gamma_2$ (the wet boundary), $\Gamma_4$ (the free surface) and possibly $\Gamma_5$ (internal free surface). We assume homogeneous and isotropic permeability. To alleviate notations, we choose units so that the specific weight of the fluid and the permeability are equal to 1. To obtain approximate fields p', we impose that $p' \geq 0$ and that conditions (4, 5, 6, 7, 8) satisfied by the true solution p are replaced by (9, 10, 11, 12, 13).

We compare two solutions differing by condition on Ω and by boundary conditions on $S_1$ and $S_3$. We assume that boundary condition on the free surface is (7). Chipot's method [4] is adapted to a little more general conditions. Any solution p' verifying (7, 9, 10, 11, 13) is a solution of the weak problem (14). According to the method of [4], one may check that this weak problem admits a solution defined as the limit of the solution of a regularised problem and that property (15) holds.

We compare now two solutions differing only by the flow condition on the free surface. We consider two pressure fields p and p' satisfying conditions (9, 10, 11, 13) with the same data and such as p satisfies (7) and p' (12). Condition (12) means that there is a loss of fluid through the free surface (evaporation). This loss of fluid is expected to lower the free surface with regard to the case where there is no loss of fluid.
One assumes first that $\Gamma_2' \cup \Gamma_4'$( concerning p') is above $\Gamma_2 \cup \Gamma_4$ (concerning p) (Fig. 2-1). One considers the new set Ω' where the part of Ω over $\Gamma_2' \cup \Gamma_4'$ has been removed. Fields p and p' satisfy the same conditions except possibly on $\Gamma_2' \cup \Gamma_4'$. On $\Gamma_2' \cup \Gamma_4'$, p and p' satisfy condition p'=0 and $-\nabla(p'+z).\underline{n} \geq 0$ . Then p and p' appear to verify the same conditions on Ω', on $\partial\Omega'$ and on free surfaces. Assuming uniqueness, we conclude that p and p' are equal on Ω' and that $\Gamma_2 \cup \Gamma_4$ and $\Gamma_2' \cup \Gamma_4'$ are merged. One can similarly argue when only a part of $\Gamma_2' \cup \Gamma_4'$ is above $\Gamma_2 \cup \Gamma_4$ (Fig. 2-2).
So $\Gamma_2' \cup \Gamma_4'$ is everywhere below $\Gamma_2 \cup \Gamma_4$ (Fig. 2-3). One is interested now in the zone Ω''' below $\Gamma_2' \cup \Gamma_4'$. On $\Gamma_2'$ we have $p \geq p' =0$, everywhere else p and p' satisfy the same conditions. Using (15), we conclude that $p \geq p'$ on Ω''' and then on Ω.
Assuming uniqueness of the solution of (14), it is then possible to assert that a field p' verifying conditions (9, 10, 11, 12, 13) is a lower bound for the true pressure p satisfying (4,5,6,7,8).

We consider now the example of a vertical dam between two reservoirs (Fig. 3). The medium obeys a Mohr-Coulomb criterion in effective stress with internal angle of friction φ and cohesion C. The very simple approximate pressure field p' (26) satisfies the conditions (9, 10, 11, 12, 13) and is so a lower bound of the true pressure field.
Using Coulomb's prism (Fig 4), we estimate the power of external forces (18) and the maximal resistant power (19). Introducing dimensionless parameters, we express kinematic condition (20, 21). The corresponding curve is given for particular values of parameters (Fig.

5). For this mechanism, the rising of fluid at the side BC is first destabilising by increasing the weight of the dam and the fluid pressure inside the medium, then, the stabilising effect of pressure on BC predominates.

## 1. Introduction

Nous cherchons à mettre en œuvre la méthode cinématique du calcul à la rupture [1,2] dans le cas d'un milieu poreux soumis à un écoulement à surface libre sans connaître la solution exacte du problème en pression. Une méthode a été proposée [3] ; elle est basée sur l'utilisation de champs de pression approchés par défaut. Pour la mettre en œuvre, il faut disposer de champs de pression approchés et être capable de vérifier qu'il s'agit bien de champs partout inférieurs ou égaux à la solution exacte sans connaître cette solution exacte. Nous proposons des conditions à vérifier par un champ de pression approché permettant de s'en assurer et ce, sous l'hypothèse d'unicité de la solution d'un problème faible en pression. Nous donnerons un exemple d'application à un barrage à parois verticales.

## 2. Rappel de la formulation cinématique utilisant un champ de pression approché [3]

On adopte la convention de la mécanique des milieux continus pour les champs de contraintes : les tractions sont comptées positivement. Un champ de pression uniforme s'écrit $-p\underline{\underline{1}}$. On suppose que le domaine de résistance peut être exprimé en contraintes effectives et que ce domaine de résistance en contraintes effectives G' vérifie la condition suivante :

$$\underline{\underline{\sigma}}' \in G' \Rightarrow (\underline{\underline{\sigma}}' - \lambda \underline{\underline{1}}) \in G', \ \forall \lambda \geq 0 \qquad (1)$$

On considère un champ de pression approché p' partout inférieur ou égal au champ de pression réel p, la fonction caractéristique $\chi$ (respectivement $\chi$') de l'ensemble p>0 (respectivement p'>0). On désigne par $\underline{\gamma}_{sat}$ (respectivement $\underline{\gamma}_d$) le poids volumique total si $\chi$=1 (respectivement ($\chi$=0). On se limite aux champs de vitesse virtuels $\underline{\hat{U}}$ tels que :

$$\underline{\gamma}_d . \underline{\hat{U}} \geq 0 \text{ si } \chi' = 0 \qquad (2)$$

Alors en notant $P_{rm}$ la puissance résistante maximale et $P_e$ la puissance des forces extérieures,

on obtient une condition nécessaire de stabilité ne faisant intervenir que le champ approché :

$$P_{rm}(p', \underline{\hat{U}}) \geq P_e(\chi', \underline{\hat{U}}) \qquad (3)$$

3. Obtention de champs de pression approchés

3.1.Formulation forte du problème de l'écoulement à surface libre

On suppose que la perméabilité est homogène et isotrope, que l'écoulement est permanent et que les forces inertielles sont négligeables. Pour alléger les notations, on suppose dans cette partie 3 que l'on choisit les unités de telle sorte que le poids volumique du fluide $\gamma_f$ et la perméabilité k (supposée homogène et isotrope) sont égaux à l'unité. Les champs approchés que nous allons considérer sont tels que p'$\geq$0 et que certaines des égalités vérifiées par la solution exacte p soient remplacées par des inégalités. Rappelons les dites conditions, en reprenant les notations de [4].

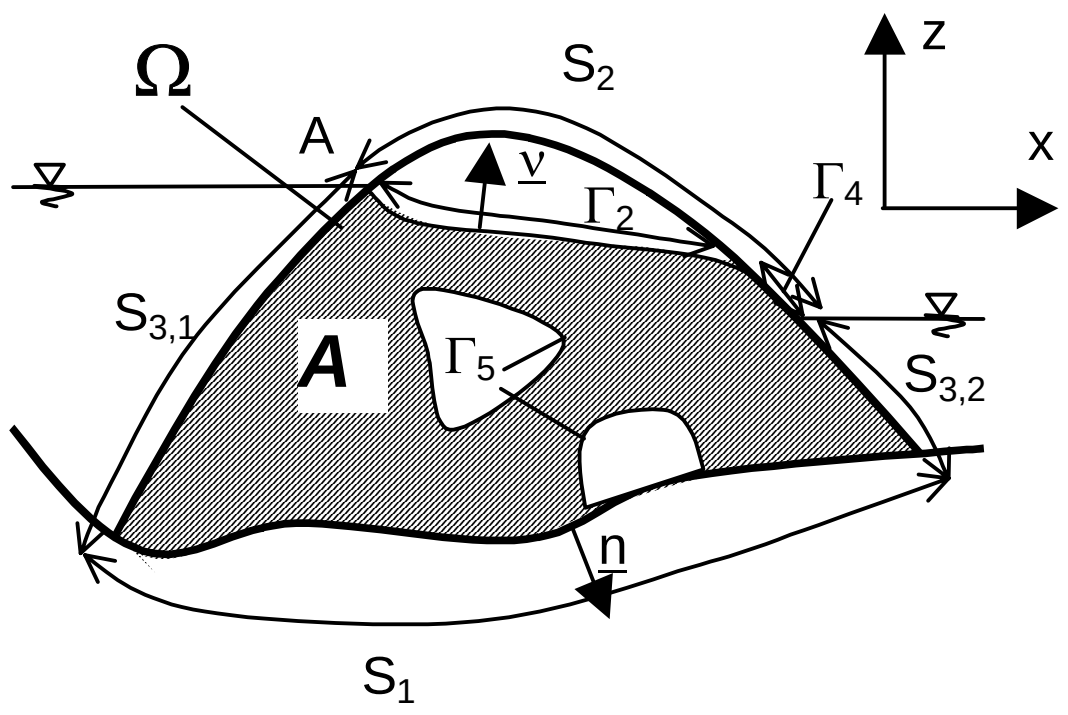

Fig. 1 Conditions aux limites pour un écoulement à surface libre dans un milieu poreux Ω

Fig.1 Boundary conditions for a free boundary flow through porous medium Ω

La frontière de Ω est décomposée en trois parties : $S_1$, $S_2$ et $S_3$ (Fig. 1). On peut envisager a priori l'apparition de surfaces libres intérieures $\Gamma_5$. La frontière de la zone saturée ***A*** (où p>0) est alors constituée de la partie de $S_1$ où p>0, de $S_3$, de la surface libre $\Gamma_2$, du bord mouillé $\Gamma_4$ et éventuellement de $\Gamma_5$. On suppose que sur une verticale, on rencontre au plus un point de $S_1$. Le débit volumique sortant $q_d$ est donné sur $S_1$, la pression $p_d$ sur $S_2 \cup S_3$.

Les conditions sur les bords de Ω et de A s'écrivent en fonction de p comme suit :

$$-\nabla(p+z).\underline{n} = q_d \text{ sur } S_1 \text{ si } p > 0 \qquad (4)$$

$p = 0, -\nabla(p+z).\underline{n} \geq 0 \text{ sur } \Gamma_4$ ( 5)

$p = p_d \geq 0 \text{ sur } S_2 \cup S_3 \text{ avec } p_d = 0 \text{ sur } S_2$ ( 6)

$p = 0, -\nabla(p+z).\underline{\nu} = 0 \text{ sur } \Gamma_2, \Gamma_5$ ( 7)

La loi de Darcy et la loi de conservation du fluide donnent :

$\Delta p = 0$ si p>0 ; p=0 sinon ( 8)

On se propose d'étudier des solutions approchées p' vérifiant les conditions suivantes:

$-\nabla(p'+z).\underline{n} = g \text{ sur } S_1 \text{ si } p' > 0 \text{ avec } g \geq q_d$ (9)

$p' = 0, -\nabla(p'+z).\underline{n} \geq 0 \text{ sur } \Gamma'_4$ (10)

$p' = p'_d \text{ sur } S_2 \cup S_3 \text{ avec } p_d \geq p'_d \geq 0 \text{ et } p'_d = 0 \text{ sur } S_2$ (11)

$p' = 0 \text{ sur } \Gamma'_2 \cup \Gamma'_5; -\nabla(p'+z).\underline{\nu} = 0 \text{ sur } \Gamma'_5; -\nabla(p'+z).\underline{\nu} = h \text{ sur } \Gamma'_2 \text{ avec } h \geq 0$ (12)

$\Delta p' = f \text{ avec } f \geq 0 \text{ si } p' > 0; p' = 0 \text{ sinon}$ (13)

3.2 Introduction d'un problème faible

Nous supposons ici que les conditions à la limite sur la surface libre et le bord mouillé sont (7). Nous reprenons avec des conditions un peu plus générales, la démarche de Chipot [4]. On introduit la fonction caractéristique $\chi$' de la partie saturée A' et $\chi'_1$ fonction caractéristique de $A' \cap S_1$. Si A' et p' sont suffisamment réguliers, alors p' satisfaisant (7, 9, 10, 11, 13) est aussi solution du problème faible suivant :

Trouver $p'$, $\chi'$, $\chi'_1$ dans $H^1(\Omega) \times L^\infty(\Omega) \times L^\infty(S_1)$ tels que :

$$\begin{cases} \text{i) } p' \geq 0 \text{ sur } \Omega, p' = p'_d \text{ sur } S_2 \cup S_3 \\ \text{ii) } 0 \leq \chi' \leq 1 \text{ sur } \Omega, \chi' = 1 \text{ p.p. sur} [p' > 0)], 0 \leq \chi'_1 \leq 1 \text{ sur } S_1, \chi'_1 = 1 \text{ p.p. sur} [p' > 0)] \\ \text{iii) } \int_\Omega (\nabla p'.\nabla\xi + \chi' \frac{\partial \xi}{\partial z} + \chi' f\xi) dV + \int_{S_1} g\chi'_1 \xi dS \leq 0 \quad \forall \xi \in H^1(\Omega), \xi = 0 \text{ sur } S_3, \xi \geq 0 \text{ sur } S_2 \end{cases} \quad (14)$$

3.3 Comparaison de deux solutions différant par la condition sur Ω et les conditions aux limites sur $\partial\Omega$

On s’intéresse à des fonctions $p_i$ satisfaisant (7, 9, 10, 11, 13) et donc solutions de (14). Suivant toujours la démarche de [4], on peut vérifier que ce problème faible (14) admet une solution obtenue comme la limite de la solution d’un problème régularisé et que ces solutions sont telles que si $p_i$ est solution de (14) avec $f_i$, $g_i$, $p_{d,i}$ alors on a la propriété suivante :

$$\{f_2 \geq f_1 \geq 0, g_2 \geq g_1 \geq 0, p_{d,1} \geq p_{d,2} \geq 0\} \Rightarrow \{p_1 \geq p_2\} \qquad (15)$$

Nous admettrons que nous sommes toujours dans des cas où la solution du problème faible est unique. Ceci n’est notamment pas vrai quand la forme d’un fond imperméable autorise la présence de bassins remplis de fluide qui ne sont pas en contact avec $S_3$ [4].

3.4 Comparaison de deux solutions différant seulement par la condition de flux sur la surface libre $\Gamma_2$

La démarche de [4] utilisée dans 3.2 et 3.3 ne semblant pas se prêter aisément à une modification de la condition sur la surface libre (7), nous traitons séparément ce cas. On suppose que l’on a deux fonctions p et p’, telles que p et p’ satisfassent les conditions (9, 10, 11, 13) avec les mêmes données, que p satisfasse la condition (7) et p’ la condition (12). La condition (12) signifie qu’il y a une perte de fluide à travers la surface libre (évaporation). Toutes choses égales par ailleurs, on s’attend physiquement à un abaissement de la surface libre par rapport au cas où il n’y a pas une telle perte de fluide.

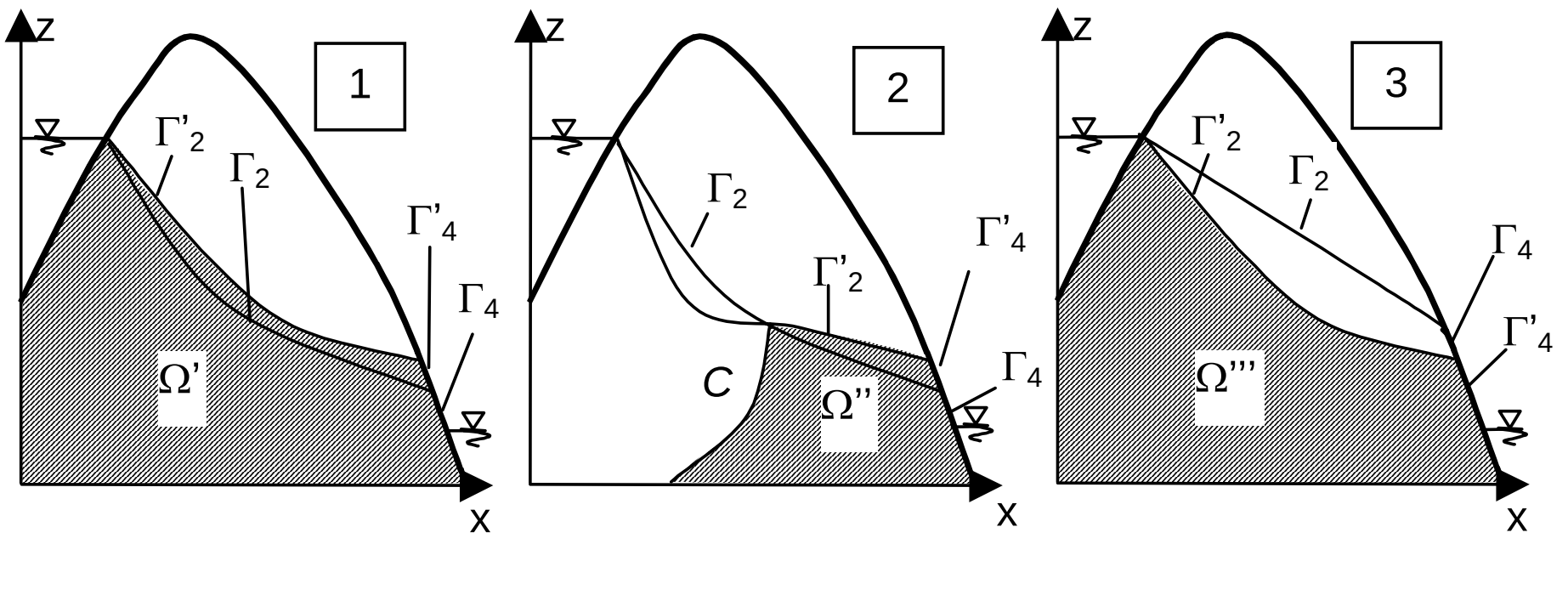

Fig. 2 Hypothèses sur la position relative de $\Gamma_2 \cup \Gamma_4$ et $\Gamma'_2 \cup \Gamma'_4$

Fig. 2 Hypotheses on relative position of $\Gamma_2 \cup \Gamma_4$ and $\Gamma'_2 \cup \Gamma'_4$

On suppose d’abord que $\Gamma'_2 \cup \Gamma'_4$ est au-dessus au sens large de $\Gamma_2 \cup \Gamma_4$ (Fig.2-1). On considère le nouveau système Ω’ (en grisé) où on enlève à Ω la partie située au-dessus de $\Gamma_2' \cup \Gamma_4'$. Sur Ω', les champs p et p' satisfont partout les mêmes conditions sauf éventuellement sur $\Gamma_2' \cup \Gamma_4'$. Sur $\Gamma_2' \cup \Gamma_4'$, p' satisfait la condition p'=0 et aussi $-\nabla(p'+z).\underline{n} \geq 0$ puisque $h \geq 0$ ; p satisfait également ces conditions. On voit qu'alors p et p' vérifient en fait les mêmes conditions sur Ω', sur ∂Ω' (incluant les bords mouillés) et sur les éventuelles surfaces libres incluses dans Ω'. Si on admet l'unicité, on a p=p' sur Ω' et on voit que $\Gamma_2' \cup \Gamma_4'$ et $\Gamma_2 \cup \Gamma_4$ sont confondues.

Si on suppose que $\Gamma_2' \cup \Gamma_4'$ n'est qu'en partie au dessus de $\Gamma_2 \cup \Gamma_4$ (Fig 2-2), on considère le sous-ensemble Ω'' de Ω limité supérieurement par la partie de $\Gamma_2' \cup \Gamma_4'$ au dessus de $\Gamma_2 \cup \Gamma_4$ et latéralement par une (ou deux) courbe *C* (cas 2D) telle que p=p' (zone hachurée de la figure 2-2). On peut appliquer alors le même raisonnement que précédemment et conclure que la partie de $\Gamma_2' \cup \Gamma_4'$ considérée est en fait incluse dans $\Gamma_2 \cup \Gamma_4$.

Donc $\Gamma_2' \cup \Gamma_4'$ est partout en dessous (au sens large) de $\Gamma_2 \cup \Gamma_4$ (Fig 2-3). On s'intéresse maintenant à la partie Ω''' de Ω située sous $\Gamma_2' \cup \Gamma_4'$. Sur $\Gamma_2'$, on a $p \geq p' = 0$ ; ailleurs les conditions vérifiées par p et p' coïncident. On en conclut en utilisant (15) que $p \geq p'$ sur Ω''' et donc aussi sur Ω puisque p' est nul sur Ω\Ω'''.

### 3.5. Conclusion

Sous réserve de l'unicité de la solution du problème faible (14), les résultats de 3.3 et de 3.4 permettent d'affirmer qu'une solution approchée vérifiant les conditions (9, 10, 11, 12, 13) est un minorant de la solution exacte. De manière analogue, on pourrait chercher des champs de pression approchés qui majorent la solution et qui sont utiles si on doit préciser les zones sèches, pour une meilleure évaluation de la puissance du poids dans le cas où certains points auraient une vitesse (virtuelle) ascendante [3].

## 4. Exemple de la stabilité d'un talus vertical compris entre deux réservoirs

### 4.1 Recherche d'une solution approchée pour la pression

On s'intéresse à un barrage vertical reposant sur un substratum horizontal imperméable (Fig. 3). Le système étudié est le rectangle de base L et de hauteur H.

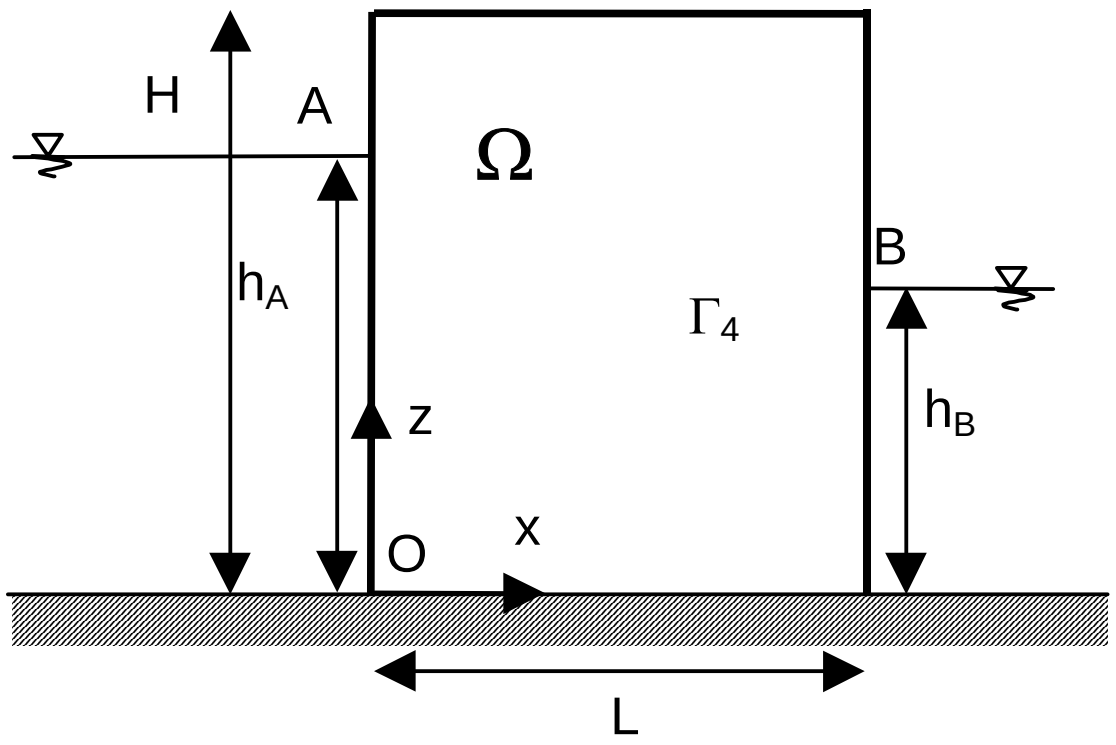

Fig. 3 Stabilité d'un barrage vertical en présence d'un écoulement à surface libre

Fig.3 Stability of a vertical dam with a free surface flow

On propose d'utiliser la solution très simple suivante en notant $\gamma_f$ le poids volumique du fluide :

$$p'(x,z) = \gamma_f \left( (h_A - z) + \frac{x}{L}(h_B - h_A) \right) \quad \text{pour} \quad z \leq h_A - \frac{x}{L}(h_A - h_B),\ p' = 0 \quad \text{sinon} \qquad (16)$$

Ce champ p' satisfait en fait les conditions à la limite de la solution exacte sur le fond imperméable et sur les bords où la pression est imposée et strictement positive ; il satisfait aussi la condition sur le laplacien (7). Ce n'est que sur la surface libre que ce champ ne satisfait pas les conditions vérifiées par la solution exacte. On vérifie qu'il satisfait (12) avec $h = ((h_A - h_B)/L)^2 \geq 0$.

4.2 Etude cinématique d'un cas particulier

On se propose d'étudier le cas particulier suivant : le matériau obéit à un critère en contraintes effectives de Mohr-Coulomb, d'angle de frottement interne φ et cohésion C ; on suppose $H=h_A$. La densité de puissance résistante maximale liée à la discontinuité de vitesse s'écrit pour ce critère de Coulomb, en notant φ l'angle de frottement interne et C la cohésion :

$$\Pi(p',[[\underline{U}]]) = (C \cot\varphi - p')[[\underline{U}]].\underline{n} \quad \text{pour} \quad [[\underline{U}]].\underline{n} \geq \left|[[\underline{U}]]\right| \sin\varphi \qquad (17)$$

Nous allons étudier la cinématique du prisme de Coulomb : le bloc de droite se détache avec une vitesse U faisant un angle β avec la surface de discontinuité. En tous les points, la vitesse a une composante verticale négative ou nulle si $\alpha + \beta \leq \pi/2$ et la condition (2) est vérifiée. Une première optimisation du mécanisme conduit à retenir $\beta = \varphi$ (Fig. 4).

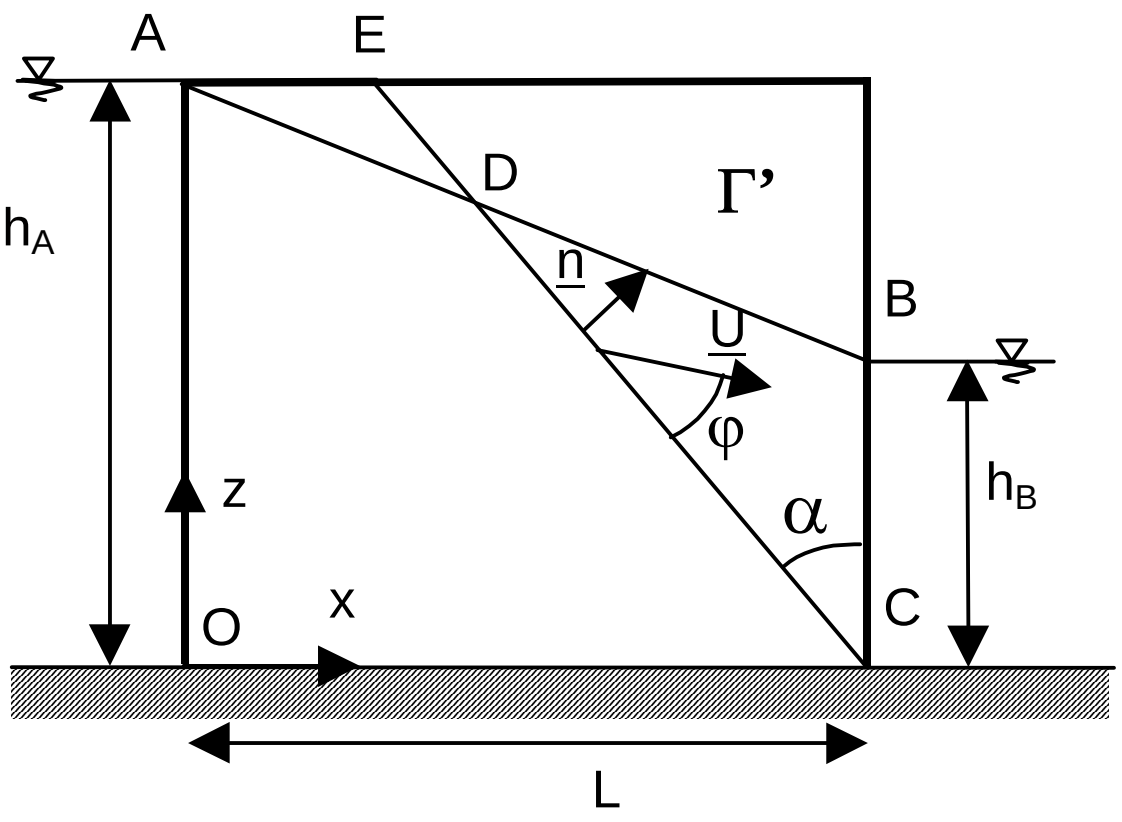

Fig. 4 Etude d'une cinématique par bloc

Fig. 4 Study of a block kinematic field

On évalue alors la puissance des forces extérieures ; il faut prendre en compte ici le poids de la zone saturée et de la zone sèche et les forces dues à la pression extérieure sur le bord BC :

$$P_e = \frac{U}{2}\left[\cos(\alpha+\varphi)\left(\frac{h_B^2}{(\cot\alpha-(h_A-h_B)/L)}\gamma_{sat} + (h_A^2\tan\alpha - \frac{h_B^2}{(\cot\alpha-(h_A-h_B)/L)})\gamma_d\right) - \sin(\alpha+\varphi)h_B^2\gamma_f\right] \qquad (18)$$

On suppose que $L \geq h_A \tan\alpha$. On évalue ensuite la puissance résistante maximale :

$$P_{rm} = \int \Pi(p',[\![U]\!])dl = \int_{CE} C\,\mathrm{cotg}\varphi\, U\sin\varphi\, dl - \int_{CD} p' U \sin\varphi\, dl = \frac{U h_A \sin\varphi}{\cos\alpha} C\,\mathrm{cotg}\varphi - \frac{1}{2}\gamma_f U \sin\varphi \frac{1}{\sin\alpha}\frac{h_B^2}{\cot\alpha-(h_A-h_B)/L} \qquad (19)$$

On note que l'effet de la pression interstitielle p' le long de CD est prise en compte dans $P_{rm}$ et non pas dans $P_e$ : la pression interstitielle le long de CD est une force intérieure au système poreux considéré qui diminue la puissance résistante maximale. On introduit les grandeurs adimensionnelles $h_B^* = h_B / h_A$ ; $L^* = L / h_A$ ; $\gamma_f^* = \gamma_f / \gamma_d$ ; $\gamma_{sat}^* = \gamma_{sat} / \gamma_d$. La condition cinématique s'écrit :

$$\frac{\gamma_d h_a}{C} \leq \frac{4\tan(\pi/4+\varphi/2)}{F(h_B^*, L^*, \gamma_f^*, \gamma_{sat}^*, \varphi, \alpha)} \qquad (20)$$

avec :

$$F = \frac{2\tan(\pi/4+\varphi/2)}{\cos(\varphi)}\left[(h_B^*)^2\gamma_f^*\left(\frac{\sin(\varphi)}{1-\tan(\alpha)\frac{(1-h_B^*)}{L^*}} - \cos(\alpha)\sin(\alpha+\varphi)\right) + (\gamma_{sat}^*-1)(h_B^*)^2\frac{\sin(\alpha)\cos(\alpha+\varphi)}{1-\tan(\alpha)\frac{(1-h_B^*)}{L^*}} + \sin(\alpha)\cos(\alpha+\varphi)\right] \quad (21)$$

Si $h_B^*$ =0 ou si $\gamma_f^*$ =0 et $\gamma_{sat}^* = 1$, seul le terme $\sin(\alpha)\cos(\alpha+\varphi)$ subsiste dans le crochet de la formule ci-dessus. On retrouve alors le résultat obtenu pour un talus vertical en l'absence de fluide [1]. Si on a $h_B^*$ =1, on peut vérifier qu'on retombe aussi sur ce même résultat en prenant cette fois comme poids volumique le poids déjaugé $\gamma_{sat}^* - \gamma_f^*$.

On optimise le choix du champ cinématique en fonction de $\alpha$ et on pose maintenant :

$$R(h_B^*, L^*, \gamma_f^*, \gamma_{sat}^*, \varphi) = \inf\left\{1/F(h_B^*, L^*, \gamma_f^*, \gamma_{sat}^*, \varphi, \alpha), \alpha \in [0, \inf(\arctan(L/h_A), \pi/2-\varphi)]\right\} \quad (22)$$

La figure 5 donne les valeurs de R pour un choix particulier des paramètres.

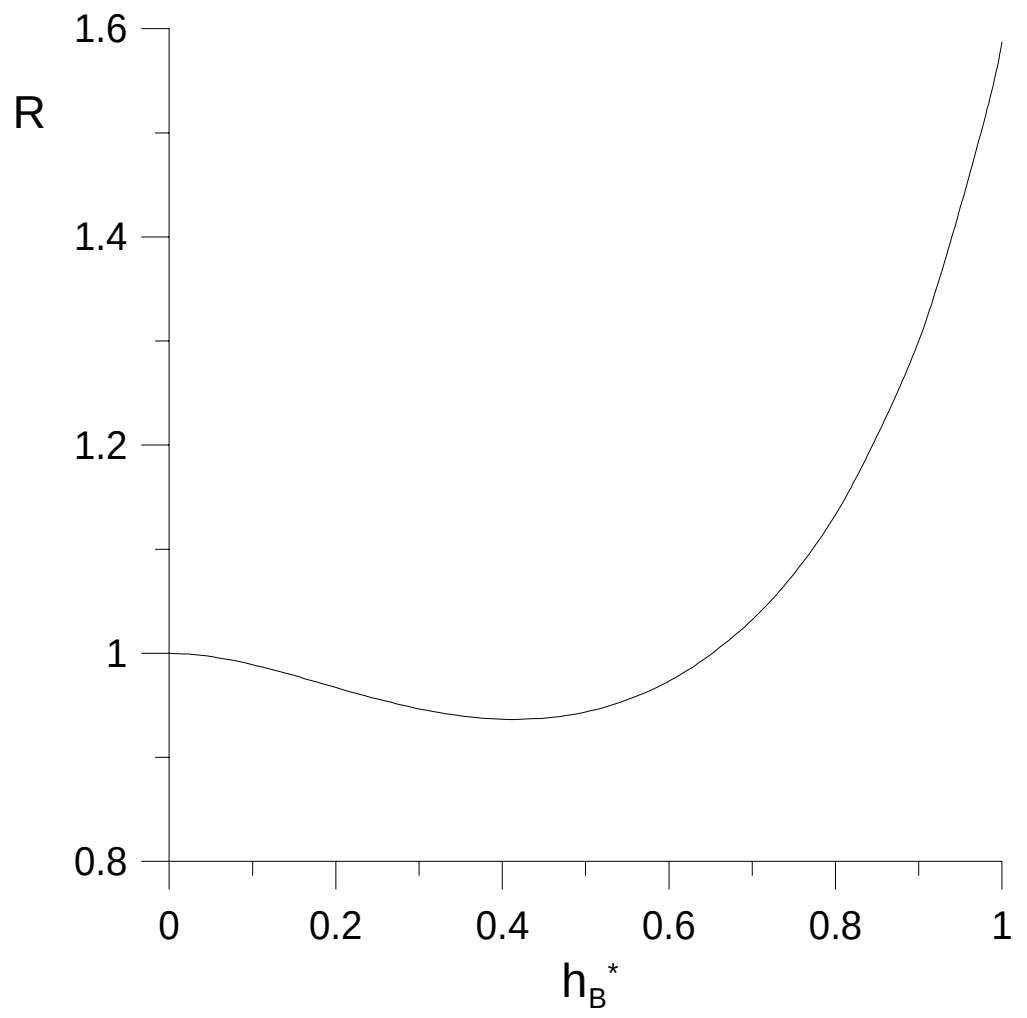

Fig. 5 Valeur de R en fonction de la hauteur d'eau $h_B^*$ ($L^* = 1$, $\gamma_f^* = 0{,}57$, $\gamma_{sat}^* = 1{,}2$, $\varphi$=30°)

Fig. 5 Value of R as a function of water height $h_B^*$ ($L^* = 1$, $\gamma_f^* = 0.57$, $\gamma_{sat}^* = 1.2$, $\varphi$=30°)

Pour ce mécanisme, l'élévation du niveau de fluide du côté BC est d'abord déstabilisant en augmentant le poids du massif et la pression du fluide au sein du massif, puis l'effet stabilisateur de la pression exercée par le fluide sur BC devient prépondérant.

## 5. Conclusion

On peut utiliser la méthode cinématique pour un milieu soumis à un écoulement à surface libre sans connaître la solution exacte du problème en pression grâce à des champs de pression approchés par défaut. Dans le cas où la perméabilité est homogène et isotrope, nous avons déterminé, sous réserve de l'unicité de la solution d'un problème faible associé, de vastes familles de champs de pression approchés par défaut en relâchant les conditions au limites, celle sur la surface libre ainsi que l'équation de champ. Nous avons montré sur l'exemple du barrage vertical l'application pratique de la méthode depuis la construction d'un champ de pression approché jusqu'à la mise en œuvre du calcul à la rupture proprement dit.

Références